\documentclass[twocolumn,showpacs,preprintnumbers,amsmath,amssymb]{revtex4}
\usepackage{graphicx}% Include figure files
\usepackage{dcolumn}% Align table columns on decimal point
\usepackage{bm}% bold math

\begin{document}

\preprint{}

\title{Testing Gravity with Muonium}% Force line breaks with \\

\author{K. Kirch}
\altaffiliation{klaus.kirch@psi.ch}
\affiliation{Paul Scherrer Institut (PSI), CH-5232 Villigen PSI, Switzerland}

\date{\today}% It is always \today, today,

\begin{abstract}
Recently a new technique for the production of muon ($\mu^+$) and muonium ($\mu^+e^-$)  beams
of unprecedented brightness has been proposed.
As one consequence and using a highly stable Mach-Zehnder type interferometer, 
a measurement of the gravitational acceleration $\bar{g}$ of muonium atoms
at the few percent level of precision appears feasible within 100\,days of running time.
The inertial mass of muonium is dominated by the mass of the positively charged - antimatter -  muon.
The measurement of $\bar{g}$ would be the first test of the gravitational interaction
of antimatter, of a purely leptonic system, and of particles of the second generation.
\end{abstract}

%Use showkeys class option if keyword
\maketitle

The gravitational acceleration of antimatter has not been measured so far.
An experiment with antiprotons (see~\cite{Hol04} and references therein) did not succeed
because of the extreme difficulty to sufficiently shield the interaction region
from electromagnetic fields. For a similar reason, results of 
measurements with electrons~\cite{Wit67} are discussed very controversial
and the plan to eventually compare with positrons was never realized.
Not affected by these problems are neutral systems like antihydrogen %(see, e.g.,~\cite{})
and, consequently,  considerable effort today is devoted to the preparation of suitable
samples of antihydrogen (compare~\cite{Hol04}).
A possibility to measure the effect of gravitation on neutral particles
is via a phase acquired in the gravitational potential
in a suitably built interferometer, demonstrated in the classic 
Colella--Overhauser--Werner (COW) experiment~\cite{Col75}.
In case of limited source performance, when one has to deal with
extended sources, comparatively large beam divergence and poor energy definition,
Mach-Zehnder type interferometers have striking advantages~\cite{Cha75}. Their
performance has been demonstrated, among others, with neutrons~\cite{Gru89} and atoms~\cite{Kei91}. 
The idea to apply interferometry to the measurement of an antimatter system
was inspired by the COW-experiments and dates back, as far as I know, to the 1980s~\cite{Sim95} 
but was put into print, with explicit mentioning of antihydrogen, positronium and 
antineutrons, first in 1997~\cite{Phi97}. 
Common problems of the species are the quality of the particle beams
and the availability of suitable, sufficiently large gratings. 
The case of positronium was further elaborated suggesting the use of
standing light waves as diffraction gratings~\cite{Obe02} but the
realization of an experiment appears still very challenging.
In the meantime also other experimental 
approaches to measure the gravitational interaction 
have been proposed for antihydrogen
and positronium, see~\cite{Hol04} for an overview.

No discussion about a gravity experiment using muonium atoms (M = $\mu^+e^-$) 
appeared in the literature yet and
the original idea of using M atoms for testing antimatter gravity
is again by Simons~\cite{Sim95}. 
The suitability of M atoms comes from the fact that the inertial mass of the muon is some
207 times larger than the one of the electron, thus,
muonium  is almost completey, to 99.5\%, antimatter.
An interesting feature is that M atoms are almost exclusively produced
at thermal energies by stopping $\mu^+$ in matter which they often leave again
as thermalized, hydrogen-like, M atom.
However, up to until recently, a gravitational experiment with muonium
would have been science fiction.
The reason for this publication is, that there is now the real chance to perform
such an experiment within the next few years.

An experiment with M atoms would constitute the first test 
of the gravitational interaction of antimatter with matter. 
It would also be the first and probably unique test of particles
of the second generation. While it would also be the first test in
a purely leptonic system one should note that tests of the 
equivalence principle proving at a high level of precision that the gravitational
interaction is independent of composition of test masses also in principle prove
(to still impressive precision) 
that electrons fall in the same way as the rest of the material. For a recent review on
tests of the equivalence principle see~\cite{Gun05}.

As a first measurement, the determination of the sign of interaction could be already interesting
(for a discussion of antigravity see~\cite{Nie91}, but also, e.g,~\cite{Kow96}), however,
a reasonable first goal for such an experiment would be to determine $\bar{g}$ 
to better than 10\%. One should add here, that it is not at all obvious that
there could be a discrepancy between the gravitational interaction of matter and antimatter,
see~\cite{Ade91}. But the universality of~\cite{Ade91} has been disputed and possible scenarios
have been sketched in~\cite{Nie91}. Anyhow, an experimentalist will probably favor the
direct measurement (and this, again, not only with respect to antimatter but also to a lepton
of the second generation) over the discussion of models.
The following quote from~\cite{Nie91} for {\it antiprotons} 
holds equally well for {\it muonium atoms}:
``It would be the first test of gravity, i.e. general relativity, in the realm
of antimatter. Even if the experiment finds exactly what one expects,
namely that antimatter falls toward the earth just as matter does, it would be,
'A classic, one for the text books.' .... Of course, if a new effect were
found in the {\it antiproton} gravity experiment, then there would be no telling what
exciting physics could follow.''

The muonium experiment appears feasible now because of two recent inventions: (i) 
a new technique to stop, extract and compress a high intensity beam of positive muons,
to reaccelerate the muons to 10\,keV and focus them into a beam spot 
of  100\,$\mu$m diameter or even less~\cite{Taq06a}; 
and (ii) a new technique to efficiently convert the muons
to M atoms in superfluid helium at or below 0.5\,K in which they thermalize
and from which they get boosted by 270\,K perpendicular to the surface 
when they leave into vacuum~\cite{Taq06b}.

Assuming an existing surface muon beam of  highest intensity as input, see e.g.~\cite{Pro06},
it should be possible to obtain an almost monochromatic beam of M atoms 
($\Delta E / E \approx 0.5/270$) with a velocity of about 6300\,m/s 
(corresponding to 270\,K or a wavelength $\lambda\approx5.6{\rm \AA}$)
and a 1-dimensional divergence of $\sqrt{\Delta E / E} \approx 43$\,mrad
at a rate of about $10^5$\,s$^{-1}$ M atoms~\cite{Taq06b}.
This is a many orders of magnitude brighter beam than available up to now.

\begin{figure}[t]
\includegraphics[width=1.\linewidth, angle=0]{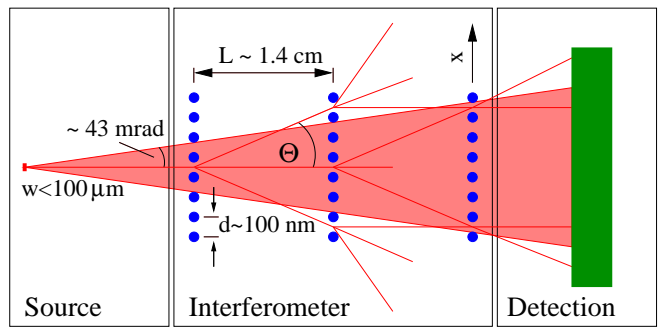}
\caption{\label{setup.eps}
Scheme of the experimental setup: 
the M beam comes from the cryogenic $\mu^+$ beam target on the left hand side, enters and 
partially traverses the interferometer and reaches the detection region on the right hand side.
The dimensions are not to scale and the diffraction angle $\theta$ is in reality smaller than the divergence.}
\end{figure}

Following the approach of~\cite{Gru89,Kei91,Phi97,Obe02} a Mach-Zehnder type interferometer
should be used in the muonium experiment.
The principle with the source, the three grating interferometer and the detection
region is sketched in Fig.~\ref{setup.eps}. We assume here three identical gratings
and use the first two for setting up the interference pattern which is scanned by moving
the third grating.
The setup is rather short, because the decay length of the
M atoms is about 1.4\,cm only ($\tau_\mu=2.2\,\mu$s).
The whole system from source to detection may be 4 decay lengths long,
and without further collimation the source illuminates a cross section of less than 5\,mm over the
length of the interferometer. The three free-standing gratings can be made sufficiently large 
with existing, proven technology with a period of 100\,nm~\cite{Sav96,Gri00}
resulting in a diffraction angle $\theta=\lambda/d\approx5.6$\,mrad.
The optimum distance $L$ between two gratings is slightly larger than one decay length; 
however, for simplicity here $L=1.4$\,cm.
Assuming another length $L$ each, for distances of the source
and the detector to the nearest interferometer grating, results in 
4 decay lenghts. Decay and transmission loss by the three  50\% open ratio gratings
reduces the initial M rate by a factor $2\times10^{-3}$, yielding $N_0=200$\,s$^{-1}$ detected M.
Because only the indicated first order diffraction carries the desired information but 
essentially all transmitted M are detected,
the interference pattern has a reduced contrast of somewhat below 4/9.
Assuming a contrast of $C=0.3$ and using eqn. (3) of~\cite{Obe02} yields
the statistical sensitivity of the experiment:
\begin{eqnarray}
S&=&\frac{1}{C \sqrt{N_0}}\frac{d}{2\pi}\frac{1}{\tau^2}\\
&\approx& 0.3\,{\rm g} \,\,\, {\rm per \,\sqrt{\#days}} 
\end{eqnarray}
which means that the sign of $\bar{g}$ is fixed after one day and
3\% accuracy can be achieved after 100\,days of running.

With the quite satisfactory statistics,
the next important  issues are the alignment and stability
of the interferometer. 
The gravitational phase shift to be observed is 
(using the notation of~\cite{Obe02})
\begin{equation}
\Phi_{\rm g}=\frac{2\pi}{d}\,{\rm g} \, \tau^2\approx0.003.
\end{equation}
This is rather small but still an order of magnitude larger than
the phase shift due to the acceleration induced by 
the rotation of the earth 
(Sagnac effect: $4\pi \tau^2 v/d \times \omega_{\rm earth}\approx 3\times 10^{-4}$). 
Other accelerations of the
system as a whole, e.g. from environmental noise, mainly
affect the contrast and must therefore be suppressed.
The same is true for misalignments of the gratings and their drifts.
The effects must be kept below the phase shift,
for example, for an unwanted translation $\Delta x$ 
of the third (scanning) grating perpendicular
to the M beam and the lines of the grating
one requires
\begin{equation}
2\pi \frac{\Delta x}{d} \le \Phi_{\rm g}
\end{equation}
and consequently 
\begin{equation}
\Delta x < 0.5\,{\rm \AA}= 50\,{\rm pm}.
\end{equation}
Rotational misalignment of the gratings around the M beam must be much less than the
period over beam height ratio, 100\,nm/5\,mm, or 20\,$\mu$rad and corresponding
drifts must not exceed 20\,nrad. In a similar way, limits for all other
static or dynamic deviations from the perfect alignment of the three identical,
equidistant, parallel gratings can be obtained.

The relatively small size of the interferometer is a major advantage 
for the stabilization.
As in previous matter interferometry experiments~\cite{Gru89,Kei91} 
the muonium experiment must use (multiple) laser interferometry
for alignment, monitoring and feedback position stabilization. The gratings
for the laser interferometry are ideally integrated in the M atom gratings
as perfect alignment is required.
State of the art piezo systems can be used for positioning the gratings and
for scanning of the third grating with a step precision of 50\,pm or better.
Because it will not be possible to perform a comparative experiment with the
($\mu^-e^+$) system, the gravitational effect on the M atoms will be 
calibrated with a laser interferometer: In case of a vertical separation
of the beam paths in the inteferometer, as in Fig.~\ref{setup.eps}, the M fringes
will be shifted by gravity with respect to the laser interference pattern. 
By a $\pi/2$ rotation of the  setup around the beam axis the gravity shift can be removed 
from the M interference while the laser stays unaffected 
(the gravitational redshift effects can be safely
neglected at the level of precision required here). Of course, the Sagnac effect has to be
taken into account, which does not affect the interference in the vertical 
but will do so in the horizontal position. Given the possible sensitivity of the
experiment it might be just at the edge, but in principle can be
measured by horizontal measurements at $\pi/2$ and
$-\pi/2$ rotations of the setup, respectively. 
Also a setup with a vertical M beam would allow studying the Sagnac effect
and the stability of the interferometer. 
This could also be a first stage of the experiment because for practical reasons
the M beam development would probably first produce a vertical beam~\cite{Taq06b}.

The muonium detection itself is straight forward: the $e^+$ from the $\mu^+$ decay
must be detected in a way that the position of the M atom can be determined to be
behind the interferometer. A coincidence signal from the detection system is desirable
in order to suppress background coming from the 3 orders of magnitude more muon decays
upstream of the third grating.
One way to achieve this is to detect both, the decay $e^+$ and the remaining atomic $e^-$ from  
the M atom decay in flight. The $e^+$ (energies up to 53\,MeV) 
could be detected in a scintillation detector which is segmented 
along the beam axis and surrounds the M beam. 
The $e^-$ (energies of 10\,eV) will be electrostatically accelerated out of the decay
region onto, e.g., a microchannel plate.

In conclusion, a muonium gravity experiment appears feasible today. The development
of the required M beam could be done within a few years and the interferometer
set up in parallel and tested with its lasers interferometry offline.
It appears possible to have the gravitational acceleration of M atoms
measured to better than 10\% within the next 5 years.

\section*{Acknowledgements}
I am indebted to L.~M.~Simons who, inspired by the Colella-Overhauser-Werner
experiments,  put forward the idea of measuring
the gravitational interaction of antimatter with a Mach-Zehnder interferometer
at a conference in the 1980s. He has also repeatedly argued that muonium
would be almost completely antimatter and thus be a reasonable target for
an experimental test of its gravitational interaction - suffering only from
insufficient source strength. 
I am also thankful to D.~Taqqu for discussions and explanations concerning 
the production of a suitable muonium beam.

\end{document}